# Optical detection of strain and doping inhomogenieties in single layer MoS$_2$

A. Michail[1,2], N. Delikoukos[1,2], J. Parthenios[2,*], C. Galiotis[2,3] and K. Papagelis[2,4,*]

[1]Department of Physics, University of Patras, Patras, 26504, Greece

[2]FORTH/ ICE-HT, Stadiou str Platani, 26504, Patras, Greece

[3]Department of Chemical Engineering, University of Patras, Patras, 26504, Greece

[4]Department of Materials Science, University of Patras, Patras, 26504, Greece

*Authors to whom correspondence should be addressed. Electronic addresses: jparthen@iceht.forth.gr and kpapag@upatras.gr.

**Abstract**: Van der Waals single-layer materials are characterized by an inherent extremely low bending rigidity and therefore are prone to nanoscale structural modifications due to substrate interactions. Such interactions can induce excess charge concentration, conformational ripples and residual mechanical strain. In this work, we employed spatially resolved Raman and Photoluminescence images to investigate strain and doping inhomogeneities in a single layer exfoliated Molybdenum disulphide crystal. We have found that correlations between the spectral parameters of the most prominent Raman bands A1´ and E´ enable us to decouple and quantify strain and charge doping effects. In comparison with AFM topography, we show that the spatial distribution of the linewidth of the A-exciton peak is strain sensitive and can capture features smaller than the laser spot size. The presented optical analysis may have implications in the development of high-quality devices based on two-dimensional materials since structural and electronic modifications affect considerably their carrier mobility and conductivity.

Since the discovery of graphene in 2004[1], the broad family of van der Waals layered materials have attracted intense scientific attention due to their unique properties[2] and expected impact in several technological fields[3]. Molybdenum disulfide ($MoS_2$), a well-known semiconductor with an indirect electronic band gap of 1.3eV[4], is a promising optoelectronic material since its single layer counterpart (SL) is transformed into semiconductor with a direct electronic gap of 1.9eV.[5,6] The intrinsic direct bandgap, its high in-plane mobility (greater than 200cm$^2$V$^{-1}$s$^{-1}$) and high on-off current ratio (~10$^8$) makes $MoS_2$ a strong competitor to graphene for many applications and especially in low power digital electronics.[7] Moreover, the outstanding mechanical properties of single-layer (SL) $MoS_2$[8] in combination with the huge band gap strain sensitivity (-45meV/% for uniaxial strain up to 2%)[9,10], the rich direct-to-indirect (~2%)[9] and semiconducting-to-metallic (~10%)[10] electronic transitions upon strain, offers a degree of freedom for the exploitation of the electronic properties of SL $MoS_2$ using strain engineering strategies.

With respect to applications in $MoS_2$ devices the adhesion and interfacial properties between the layered dichalcogenide and the supporting substrate is of great importance. It is well documented for graphene that crystalline monolayer materials experience uncontrollable mechanical fields via the preparation (e.g. exfoliation[11] or chemical vapor deposition[12]) or the transfer processes along with charge doping mainly induced by the interaction of the material with the substrate.[13] These residual mechanical fields affect considerably the electronic properties and, therefore, the quantification of strain and doping levels in $MoS_2$ is a prerequisite for its proper characterization and implementation in nanoscale devices.

Recently, it has been found that the characteristic Raman modes of $MoS_2$ are affected by mechanical strain[9,14-17] and changes in charge carrier concentration.[18,19] The degeneracy of $E'$ mode is lifted for uniaxial strains higher than 1%, leading to a band splitting.[9,15] A number of studies concerning the effect of uniaxial strain on $MoS_2$ report that the Grüneisen parameter -the change of the mode frequency with respect to strain- of the $E'$ vibration lies between 0.64 and 1.10[9,14,15]. The Grüneisen parameter of the out-of- plane $A_1'$ vibration is around 0.21[14], although some existing works report the absence of frequency shifts for this mode upon mechanical deformation.[9,15] In Table S1 the available experimental results regarding the application of uniaxial strain on exfoliated monolayers of $MoS_2$ are summarized. Electronic doping, on the other hand, leads to the occupation of the anti-bonding states at the bottom of the conduction band located at the K-point of the Brillouin Zone, yielding significant alterations of the electron-phonon coupling (EPC) of the recorded Raman active modes.[18] The EPC of the $E'$ mode is much lower compared to the out-of-plane mode $A_1'$.[18] Since the phonon linewidth is proportional to EPC [20] the $A_1'$ linewidth increases by 6 cm$^{-1}$ while the $E'$ one does not experience any significant variation as a function of doping level.[18] For a maximum electron concentration, $n$, of 1.8x10$^{13}$ cm$^{-2}$ the shift rates with $n$ for $A_1'$ and $E'$ mode frequencies have been determined.[18] Therefore, the coexistence of residual strain and charge doping in monolayer MoS$_2$ complicates Raman characterization. A similar complication exists in graphene and a method to decouple both contributions by means of Raman spectroscopy has been developed previously.[21,22] In this work, we have performed Raman and photoluminescence mapping (25 spectra per 1 μm$^2$) in mechanically exfoliated monolayer MoS$_2$. The correlation between the previously extracted Raman and PL spectral parameters can provide, among others, important information about the presence and the degree of residual strain and unintentional doping. Moreover, it is demonstrated the notable sensitivity of both Raman and PL spectroscopy on length-scales below the laser spot by direct comparison with AFM topography.

For this purpose, a monolayer MoS$_2$ crystal was exfoliated from a bulk natural MoS$_2$ crystal (SPI Supplies) and deposited on SiO$_2$ (300nm)/Si substrate by micromechanical exfoliation. Optical microscopy (Figure 1(a)) and Raman spectroscopy Figure 1(b) were used to identify the monolayer. A detailed Raman mapping, with a 200nm step-size was conducted. In total, ~ 2900 spectra were collected in the backscattering geometry using a Renishaw InVia2000 Raman spectrometer equipped with a 2400groves/mm grating and providing ~2cm$^{-1}$ spectral resolution. The 514.5nm line of an Ar$^+$ laser was used for excitation, focused on the sample by means of x100 (NA=0.85) objective at a power lower than 0.7mW. The laser spot size was ~590 nm (see supplemental material details for the spot size determination). It is interesting to point out that the Full Width at Half Maximum (FWHM) of both Raman bands $A_1'$, $E'$ is comparable to spectrometer's response and it is necessary to decouple the spectrometer response function from the recorded spectral widths. The observed lineshapes exhibit a Voigt spectral profile corresponding to a convolution of the intrinsic phonon linewidth (Lorentzian component) and the instrumental response function (Gaussian component).[23] The linewidths of Lorentzian components for the $A_1'$ and $E'$ bands were extracted by fitting Voigt profiles to the available experimental spectra, fixing the width of the Gaussian component to 1.9cm$^{-1}$ as determined by the FWHM of neon lamp spectral lines. This type of spectra analysis should be applied not only in MoS$_2$ but in every 2D material (e.g. BN, WS$_2$, MoSe$_2$), where the FWHM of the Raman bands is of the order of the spectrograph's response.

Figure 1(b) presents a Raman image obtained from the monolayer; the color scale renders the $\Delta\omega$ values, namely the frequency difference between the $A_1'$ and $E'$ modes. The mean $\Delta\omega$ is 18.3cm$^{-1}$ indicating a single layer crystal.[24] The edges of the monolayer are found to exhibit systematically higher Δω values (~20cm$^{-1}$), while in the interior of the crystal well defined regions exist showing slightly lower Δω values around 17.5cm$^{-1}$. Important information can be extracted by means of the correlation plot FWHM($A_1'$) *vs* Pos($A_1'$) presented in Figure 1(c). It is evident that the experimental points form clearly established clusters spatially correlated with specific sample regions (inset of Figure 1(c)). In particular, the region of black squares originates from the thicker areas of the sample. The left and right edges of the SL MoS$_2$ give rise to two different clusters shown as orange diamonds and blue inverted triangles, respectively. The green triangles stem from measurements taken at the borderline between bulk and monolayer MoS$_2$. More information about Raman spectra as well as the origin and symmetry of the main Raman modes are given in the supplemental. The evolution of the Raman spectra as the laser beam crosses the aforementioned boundary is presented in Figure S1 and Figure S2.

The group of points represented as dark gray crosses exhibit systematically higher FWHM($A_1'$). A Raman map of the sample by means of the FWHM($A_1'$) is shown in Figure S3. These points originated from relatively small size interior sample regions, depicted as grey in the inset of Figure 1(c). These regions are nicely correlated with areas in the AFM topography (Figure 1(d)), marked with white circles, attributed to conformational features such as nanometer scale wrinkling, bubbles or molecular trapping. Similar features have been recently detected by Raman spectroscopy in graphene[22] by means of the broadening of the Raman 2D band. Interestingly, the same behavior is observed in the photoluminescence (PL) linewidth map (*vide infra*). The comparison between AFM topography and Raman maps reveals the extreme sensitivity of the later method of being capable to detect features at the submicron level. For instance, the feature indicated by the white arrow in Figure 1(d) has a diameter of about 250 nm, is clearly resolved in the Raman map (see inset of Figure 1(a)). Similar spatial sensitivity is achieved by the linewidth of the PL peak (see below).

The remaining red circular points in Figure1(c) come from the interior of the flake and spread around a straight line with a slope of -1.6±0.1 (see Figure S5), verifying the presence of inhomogeneous doping in the SL MoS$_2$ mainly due to the interaction with the underlying substrate. Analogous information can be extracted from the Pos($A_1'$) vs Pos($E'$) correlation plot (Figure S6) but the aforementioned groups of points cannot be clearly distinguished, thus hampering the very concise spatial correlation shown in Figure 1(c). From this point onward, the discussion will be focused solely on the interior points of the sample, excluding the edges and the features (dark grey crosses in Figure 1(c)) mentioned above.

In the following we have undertaken a systematic study to separate by optical means the effects of strain and unitentional doping in the interior of the exfoliated MoS$_2$ monlayer. During the deposition and peeling procedures it is assumed that the flake is subjected to biaxial strain[11,21,22]. Therefore, the strain induced phonon frequency shift, $\delta\omega$, can be estimated by the relation $\delta\omega = 2\gamma\omega_o\varepsilon$, where $\omega_o$ is the frequency of the particular mode at zero strain, $\gamma$ is its Grüneisen parameter and $\varepsilon$ is the applied biaxial strain.[25] Moreover, calibration curves for the $A_1'$ and $E'$ phonon frequecncies as a function of doping level have been establised.[18] It is important to stress that an unstrained/undoped sample is an unrealistic situation. A simply-supported MoS$_2$ flake is normally subjected to residual stresses resulting from the transfer procedure or the roughness of the underlying substrate along with substrate induced doping.[11] Therefore, the unperturbed phonon frequency is difficult to be obtained experimentally. However, the mean value of the recorded frequencies in a detailed Raman mapping all over the area of the examined flake can be used as a reference. In this case, the presented analysis can provide information on strain and charge doping level relative to the corresponding mean "state" of the spectroscopically probed area. As presented in more details in supplemental material, strain and charge carrier concentration can be expressed linearly in terms of the Pos($A_1'$), Pos($E'$), assuming resonably that at low levels of strain and charge doping both effects are decoupled form each other. In other words, a linear transformation between the vector spaces $\Delta Pos(A_1')$-$\Delta Pos(E')$ and $\varepsilon$-$n$ takes place that allow us to quantify the strain and doping levels within the $Pos(A_1')$-$Pos(E')$ plot presented in Figure 2(a).

In order to define the $\varepsilon$-$n$ axes a Grüneisen parameter value of 0.86 (average value of Table S1) for the $E'$ mode was used, while a Grüneisen parameter of 0.15 for the $A_1'$ mode was chosen.[14] The shift of the $A_1'$ and $E'$ modes with electron concentration are about -4cm$^{-1}$ and -0.6cm$^{-1}$ for a maximum electron concentration of 1.8x10$^{13}$ cm$^{-2}$, respectively.[18] As mentioned above, since the zero strain and charge neutrality phonon frequencies are unknown, the origin of the $\varepsilon$-$n$ axes is considered at the mean frequencies of the $A_1'$ and $E'$ modes located at point ( 384.6±0.2 cm$^{-1}$, 402.7±0.2 cm$^{-1}$). The calculated Raman strain and doping maps are presented in Figure 2(b) and Figure S8(a), respectively. The maximum strain and doping differences are of the order of 0.08% and 0.5x10$^{13}$cm$^{-2}$, respectively (Figure S8(b,c)). It is important to note that as long as the zero strain and charge neutrality points are undefined, the type of strain (tensile or compressive) and charge carrier (electron or hole) cannot be determined.

Complementary results about the residual strain metrology in SL MoS$_2$ can be acquired using PL spectroscopy. The PL mapping was conducted using a diffraction grating of 1200 groves/mm while keeping all the experimental conditions, including the probed sample region, identical to the Raman measurements. A characteristic PL spectrum of SL MoS$_2$ is presented in Figure S9. In the spectral range of 1.7eV to 1.95eV a single photoluminescence peak is observed attributed to the direct $A$-exciton transition centered at around 1.84eV.[5,26] In Figure S10 (a) a map of PL peak intensity is presented. The PL signal in SL MoS$_2$ is much stronger than the Raman signal[26], and as a result PL spectra of mesurable intensity were aquired at positions 0.8μm away from the crystal edges (Figure S10(b)). As can be clearly seen, the PL mapping overestimates the spatial extend of the crystal and does not allow direct comparison with the Raman results. As explained in the supplemental material, the usage of the Raman signal from the underlying Si substrate allows us to define accuratlely the borders of the MoS$_2$ flake. In Figure 3(a), the PL linewidth map is depicted, exhibiting a variation of about 100meV in the monoalyer area. The conformational features which were identified in the Raman and AFM measurements, are also clearly captured by the PL FWHM (Figure 3(a)) and intensity (Figure S10(a)) map. At these positions a considerable broadening and intensity reduction of the PL emisssion band is observed. The smallest optically detectable feature is indicated by the white arrow in Figure 3(a).

It is well documented that the direct optical transition of $MoS_2$ varies significantly with strain.[9,14,27,28,29] In order to estimate the strain distribution along the sample, the band gap deformation potential, namely the change in the direct optical gap per unit strain, for biaxial strain is needed. To our knowledge no experimental data are available on this physical quantity. On the other hand, a plethora of ab-initio calculations exist in the literature, that enables an estimation of the deformation potential of about -0.2eV/%.[16,28,30] This strain dependence of the direct optical band gap of $MoS_2$ was used to further verify the local strain variation relative to a reference value of 1.84eV obtained by averaging the corresponding $A$-exciton peak positions of the interior points.

In Figure 3(b) a PL-derived strain map is shown. The maximum strain range is 0.06% (Figure S11). It is important to note here that at the flake edges, the peak of the $A$-exciton blue shifts by ~50 meV relative to the PL emission peak in the interior (Figures S12(a) and S12(b)). According to both Raman (Figure 2(b)) and PL strain maps (Figure 3(b)), the strain at the edges is higher than the interior points. Gutiérrez et al has shown an opposite trend of the A-exciton peak in synthesized monolayer of $WS_2$.[31] The PL and Raman strain maps correlate nicely while some minor differences can be merely explained on the basis of the exciton funneling[27]. According to this phenomenon the photoexcited carriers drift of about 1 μm towards regions of high tensile strain before recombination.[27,32] Since the electronic band gap decreases as a function of strain, the PL is expected to provide the minimum bandgap in the illuminated sample area. In contrast, the Raman signal gives an average over the entire laser spot.

In conclusion, it was demonstrated that the correlation between various spectral parameters of the most prominent Raman peaks $A_1'$, $E'$ as well as $A$-exciton profile in SL $MoS_2$ can give important information about strain and doping variations. Also, it allows the detection of features on length scales below the laser spot diameter. In particular, the FWHM($A_1'$) vs Pos($A_1'$) plot translated spatially, and the map of the linewidth of the $A$-exciton peak can resolve sub-spot size (~250nm) features in SL $MoS_2$. The Pos($A_1'$) vs Pos($E'$) correlation allowed the separation of strain and doping while assuming that both effects are decoupled from each other, the level of strain and doping can be determined. It's noteworthy to stress that the linewidth of the main Raman peaks of MoS$_2$ and other 2D materials is comparable with the resolution of the Raman instruments, and thus it is necessary to employ Voigt profiles in the analysis of the Raman bands. The imaging of strain and charge carrier density inhomogeneities not only enables a more in-depth characterization of single layer MoS$_2$ devices but also enhances our ability to study the physical mechanisms behind the strain transfer as well as the interactions between 2D-MoS$_2$ and underlying substrate.

**Acknowledgements** This research has been co-financed by the European Union and Greek national funds through the programme "ARISTEIA II: GRAPHENE PHYSICS IN THE TIME DOMAIN AND APPLICATION TO 3D OPTICAL MEMORIES" implemented in the frame of the Operational Program "Education and Lifelong Learning". All authors acknowledge financial support from the Graphene FET Flagship ("Graphene Based Revolutions in ICT And Beyond"- Grant agreement no: 604391).

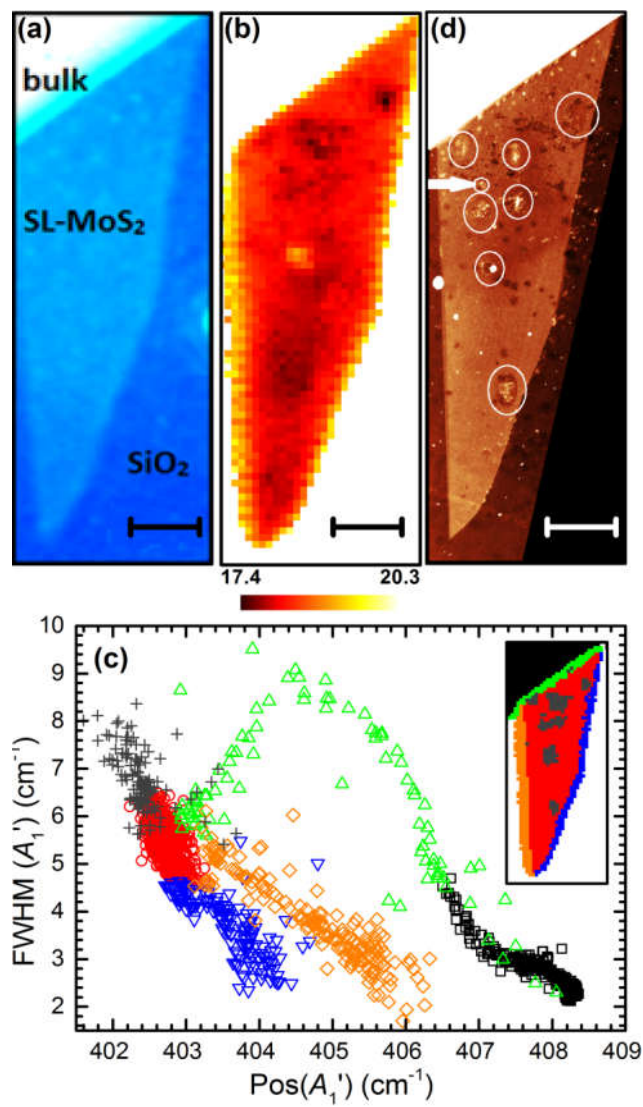

**FIGURE 1** (a) Optical image of the exfoliated monolayer MoS$_2$ deposited onto a Si/SiO$_2$ wafer (scale bar, 2μm). (b) Δω map of the monolayer area. (c) Correlation plot of FWHM($A_1'$) *vs* Pos($A_1'$). The experimental points are lying in well-defined clusters having a distinct spatial correlation depicted in the inset. (d) AFM topography of the SL-MoS$_2$ crystal. The white circles indicate the various conformational features. The white arrow points to the smallest spectroscopically detectable feature (~250nm).

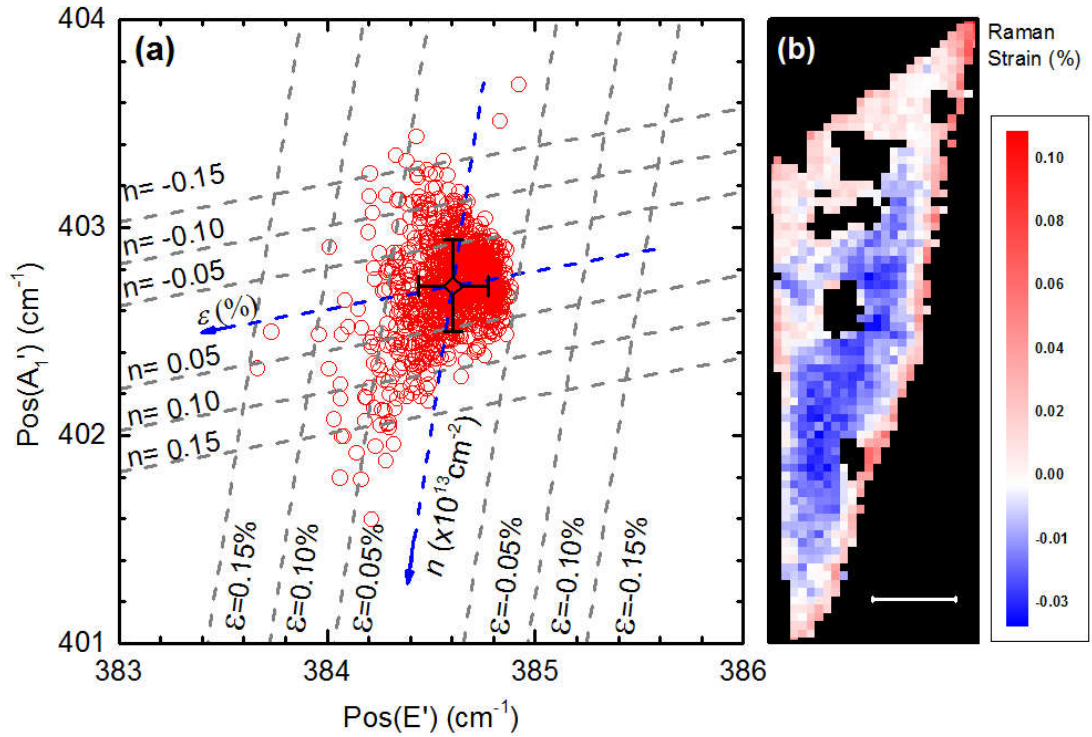

**FIGURE 2** (a) The constructed $\varepsilon$-$n$ space within the Pos($A_1'$) vs Pos($E'$) plot. Constant carrier density and biaxial strain regions are represented by dashed lines. (b) Strain map constructed from Raman measurements.

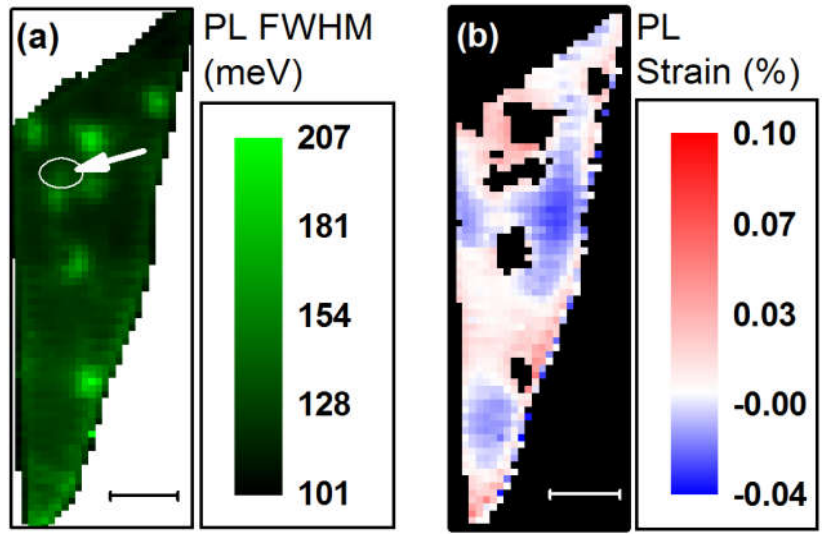

**FIGURE 3** (a) Strain map constructed from PL measurements and (b) PL linewidth map. The white arrow indicates the smallest feature (~250nm) that was optically resolved.